\documentclass{elsart}

\usepackage{latexsym}
\usepackage{natbib}
\usepackage{amsmath}
\usepackage{amssymb}
\usepackage{amsfonts}

\usepackage{ifpdf}
\usepackage{graphicx}
\ifpdf
\usepackage[%
  pdftitle={Instructions for use of the document class
    elsart},%
  pdfauthor={Simon Pepping},%
  pdfsubject={The preprint document class elsart},%
  pdfkeywords={instructions for use, elsart, document class},%
  pdfstartview=FitH,%
  bookmarks=true,%
  bookmarksopen=true,%
  breaklinks=true,%
  colorlinks=true,%
  linkcolor=blue,anchorcolor=blue,%
  citecolor=blue,filecolor=blue,%
  menucolor=blue,pagecolor=blue,%
  urlcolor=blue]{hyperref}
\else
\usepackage[%
  breaklinks=true,%
  colorlinks=true,%
  linkcolor=blue,anchorcolor=blue,%
  citecolor=blue,filecolor=blue,%
  menucolor=blue,pagecolor=blue,%
  urlcolor=blue]{hyperref}
\fi

\makeatletter
\def\elsartstyle{%
    \def\normalsize{\@setfontsize\normalsize\@xiipt{14.5}}
    \def\small{\@setfontsize\small\@xipt{13.6}}
    \let\footnotesize=\small
    \def\large{\@setfontsize\large\@xivpt{18}}
    \def\Large{\@setfontsize\Large\@xviipt{22}}
    \skip\@mpfootins = 18\p@ \@plus 2\p@
    \normalsize
}
\@ifundefined{square}{}{\let\Box\square}
\makeatother

\begin{document}

\begin{frontmatter}

\title{Second post-Newtonian approximation of scalar-tensor theory of gravity}

\author[label1]{Yi Xie}\ead{yi.s.xie@gmail.com}, 
\author[label2,label3]{Wei-Tou Ni}, \author[label2]{Peng Dong}, and \author[label1]{Tian-Yi Huang} 

\address[label1]{Department of Astronomy, Nanjing University, Nanjing, 210093 China}
\address[label2]{Center for Gravitation and Cosmology, Purple Mountain Observatory, Chinese Academy of Sciences, Nanjing, 210008 China}
\address[label3]{National Astronomical Observatories, Chinese Academy of Sciences, Beijing, 100012 China}

\begin{abstract}
	Deep space laser ranging missions like ASTROD I (Single-Spacecraft Astrodynamical Space Test of Relativity using Optical Devices) and ASTROD, together with astrometry missions like GAIA and LATOR will be able to test relativistic gravity to an unprecedented level of accuracy. More precisely, these missions will enable us to test relativistic gravity to $10^{-7}-10^{-9}$ of the size of relativistic (post-Newtonian) effects, and will require 2nd post-Newtonian approximation of relevant theories of gravity. The first post-Newtonian approximation is valid to $10^{-6}$ and the second post-Newtonian approximation is valid to $10^{-12}$ in terms of post-Newtonian effects in the solar system. The scalar-tensor theory is widely discussed and used in tests of relativistic gravity, especially after the interests in inflation models and in dark energy models. In the Lagrangian, intermediate-range gravity term has a similar form as cosmological term. Here we present the full second post-Newtonian approximation of the scalar-tensor theory including viable examples of intermediate-range gravity. We use Chandrasekhar's approach to derive the metric coefficients and the equation of the hydrodynamics governing a perfect fluid in the 2nd post-Newtonian approximation in scalar-tensor theory; all terms inclusive of $O(c^{-4})$ are retained consistently in the equations of motion.
\end{abstract}

\end{frontmatter}
\allowdisplaybreaks

\section{Introduction}
Although Einstein's general relativity has achieved great success both in experimental tests and in astrophysical applications during the last few decades, the desire to find a gravitation theory consistent with quantum theory together with the ever-increasing precision of experiments and astrophysical observations has urged many ``alternative theories'' to be proposed. Among them, the scalar-tensor theory is the most eminent one, because it is the simplest and most natural way to modify general relativity. Many modern theories, such as extra-dimensional theory, string theory, brane world and noncommutative geometry, which try to unify gravity and microscopic physics or explain the dark energy in cosmology, demand a scalar field in addition to the metric tensor. In this paper, we take a phenomenological point of view. In the low energy effective field limit, the action of most of these theories can be transformed to the following general form:
\begin{equation}
	\label{lgst}
	S=\frac{c^3}{16\pi}\int \bigg(\phi R -\frac{\theta(\phi)}{\phi}\phi^{,\sigma}\phi_{,\sigma}+2\phi\lambda(\phi)
	-\frac{16\pi}{c^4}\mathcal{L}_I(g_{\mu \nu},\phi,\Psi) \bigg)\sqrt{-g}\,\mathrm{d}^4x,
\end{equation}
where $\theta(\phi)$ and $\lambda(\phi)$ are two arbitrary functions of the scalar field $\phi$. $g=\mathrm{det}(g_{\mu \nu})<0$ is the determinant of the metric tensor $g_{\mu \nu}$, $R$ is the Ricci scalar, $\Psi$ denotes all the matter fields. Greek indexes run from $0,$ to $3$, and Latin indexes run from $1$ to $3$. The signature of $g_{\mu \nu}$ is $(-,+,+,+)$.

In general, the matter fields $\Psi$ interact with both the metric field and the scalar field. If we assume the Einstein equivalence principle to be correct, then the matter fields $\Psi$ do not interact directly with the scalar field $\phi$ and the interaction Lagrangian $\mathcal{L}_I(g_{\mu \nu},\phi,\Psi)$ becomes $\mathcal{L}_I(g_{\mu \nu},\Psi)$. Since Einstein equivalence principle is verified to a very high accuracy \citep{Ni2005}, we shall assume it is valid here. Violations of the Einstein equivalence principle have been considered in \citet{Ni2005}. Here, for simplicity, we do not consider it. Therefore, in this paper, we adopt the following action 
\begin{equation}
	\label{lgst2}
	S=\frac{c^3}{16\pi}\int \bigg(\phi R -\frac{\theta(\phi)}{\phi}\phi^{,\sigma}\phi_{,\sigma}+2\phi\lambda(\phi)
	-\frac{16\pi}{c^4}\mathcal{L}_I(g_{\mu \nu},\Psi) \bigg)\sqrt{-g}\,\mathrm{d}^4x.
\end{equation}

Depending on the functional form, the $\lambda(\phi)$ term can include (i) cosmological constant and quitessence; (ii) the mass term of scalar field which induces intermediate-range gravity. Since this paper deals with post-Newtonian approximation and assume an asymptotic flat spacetime, we include the case with the mass term and intermediate-range gravity only; we assume that
\begin{equation}
	\label{eqlambda}
	\lambda(\phi)=\lambda_2(\phi-\phi_0)^2,
\end{equation}
where $\lambda_2$ is a constant and $\phi_0$ is the constant asymptotic value of $\phi$. In this investigation, we will obtain the second post-Newtonian approximation of scalar-tensor theory of gravity including intermediate-range gravity using Eq. (\ref{lgst2}) with $\lambda(\phi)$ given by Eq. (\ref{eqlambda}).

\citet{Fujii} on the scalar-tensor theory of gravitation gives a good account of the historical development of the scalar-tensor theories. Here we present a very brief history related to our choice of Lagrangian/action Eq. (\ref{lgst2}). \citet{Jordan} first proposed scalar-tensor theory in connection with projective geometry and five-dimensional Kaluza-Klein theory \citep{Kaluza,Klein}. Assuming the validity of Einstein equivalence principle, \citet{BransDicke} reached a specification of Jordan's theory. This theory is termed Brans-Dicke-Jordan theory in the compendium of metric theories of gravity compiled in \citet{Ni1972}. This theory is a special case of (\ref{lgst2}) with $\theta(\phi)=\omega=\mathrm{const.}$ and $\lambda=0$. \citet{Bergmann1968} generalized Jordan's theory regarding to $\phi$. In his paper, the interaction Lagrangian included only electromagnetic field $F_{\mu\nu}$, not other matters, with
\begin{equation}
	\label{}
	\mathcal{L}_I=f(\phi)M,
\end{equation}
where $M$ is the Maxwell scalar formed from $F_{\mu\nu}$ and $f$ is an arbitrary function. From experiments on the test of weak equivalence principle, $f(\phi)$ is constrained as follows \citep{Ni2005}:
\begin{equation}
	\label{}
	\frac{|1-f(\phi)|}{U} < 10^{-10},
\end{equation}
where $U$ is the Newtonian potential. With $f(\phi)=1$, Bergmann's theory is of the form (\ref{lgst2}) with $\Psi$ denotes electromagnetic field $F_{\mu\nu}$. \citet{Wagoner1970} completed Bergmann's theory with other matters and postulated his principle of mutual coupling (equivalent to Einstein equivalence principle). The resulting theory is called Bergmann-Wagoner theory in the compendium of metric theories of gravity compiled in \citet{Ni1972}. Bergmann-Wagoner theory is equivalent to Eq. (\ref{lgst2}).

Deep space laser ranging missions such as ASTROD I (Single-Spacecraft Astrodynamical Space Test of Relativity using Optical Devices) and ASTROD (Astrodynamics Space Test of Relativity) \citep{ASTROD1,ASTROD2,ASTROD3}, together with astrometry missions such as Global Astrometric Interferometer for Astrophysics ({http://www.esa.int/\linebreak[3]esaSC/\linebreak[3]120377\linebreak[3]\_index\linebreak[3]\_0\_m.htm}) and Laser Astrometric Test of Relativity (LATOR) \citep{LATOR} will be able to test relativistic gravity to an unprecedented level of accuracy in the solar system. More precisely, these missions will enable us to test relativistic gravity to $10^{-7}-10^{-9}$ of the size of relativistic (post-Newtonian) effects, and will require 2nd post-Newtonian approximation of relevant theories of gravity. The first post-Newtonian approximation is valid to $10^{-6}$ and the second post-Newtonian is valid to $10^{-12}$ in terms of post-Newtonian effects in the solar system.

Here we present the full second post-Newtonian approximation of the scalar-tensor theory (\ref{lgst2}) with $\lambda(\phi)$ given by Eq. (\ref{eqlambda}) treating $\lambda_2$ as an independent parameter. We derive the metric coefficients (in Sec. \ref{sec1}); when $\lambda_2=0$, our result agrees with the result of \citet{Damour1992,Damour1996} with one scalar field where they are comparable. Damour and Esposito-Far\`{e}se use a field-theoretical point-particle approach; we use Chandrasekhar's perfect fluid approach. In Sec. \ref{sec2}, we derive the equation of the hydrodynamics governing a perfect fluid in the 2nd post-Newtonian approximation; all terms inclusive of $O(c^{-4})$ are retained consistently in the equation of motion. In Sec. \ref{sec3}, the various conserved quantities to $O(c^{-4})$ are isolated with the aid of the energy-momentum complex. In Sec. \ref{sec4}, we discuss solar-system dynamics and test of scalar-tensor theory. In Sec. \ref{sec5}, we present an outlook for further works.

\section{Second post-Newtonian approximation}
\label{sec1}
\subsection{Field equations}
Variation of the action Eq. (\ref{lgst2}) with respect to $g_{\alpha \beta}$ yields
\begin{equation}
	\label{feg}
	R_{\mu \nu}  = \frac{8\pi}{\phi c^2}\bigg(T_{\mu \nu}-\frac{1}{2}g_{\mu \nu}T\bigg)+\frac{\theta(\phi)}{\phi^2}\phi_{,\mu}\phi_{,\nu}+\frac{1}{\phi}\bigg(\phi_{;\mu\nu}+\frac{1}{2}g_{\mu \nu}\Box_g\phi\bigg)-g_{\mu\nu}\lambda,
\end{equation}
where $\Box_g(\cdot)=(\cdot)_{;\alpha \beta}g^{\alpha \beta}$. $T_{\mu \nu}$ is the stress-energy-momentum tensor of matter defined as \citep{Landau}
\begin{equation}
	\label{}
	\frac{c^2}{2}\sqrt{-g}T_{\mu\nu}\equiv\frac{\partial(\sqrt{-g}\mathcal{L}_I)}{\partial g^{\mu\nu}}-\frac{\partial}{\partial x^\alpha}\frac{\partial(\sqrt{-g}\mathcal{L}_I)}{\partial g^{\mu\nu}_{,\alpha}},
\end{equation}
and $T$ is the trace of $T^{\mu\nu}$. Here we write $T^{\mu\nu}$ in the following form
\begin{equation}
	\label{}
	c^2T^{\mu \nu}=\rho (c^2+\Pi) u^{\mu} u^{\nu}+\pi^{\mu \nu},
\end{equation}
where $\rho$ and $\Pi$ are the density and the internal energy in the co-moving frame, $u^{\mu}$ is the dimensionless 4-velocity and $\pi^{\mu\nu}$ is the anisotropic stress tensor.
For perfect fluids,
\begin{equation}
	\label{}
	\pi^{\mu \nu}=(g^{\mu \nu}+u^{\mu} u^{\nu})p,
\end{equation}
where $p$ is an isotropic pressure.

Variation of the action with respect to $\phi$ yields
\begin{equation}
	\label{fep}
	\Box_g\phi=\frac{1}{3+2\theta(\phi)}\bigg(\frac{8\pi}{c^2}T-\phi_{,\alpha}\phi^{,\alpha}\frac{\mathrm{d}\theta}{\mathrm{d}\phi}-2\phi^2\frac{\mathrm{d}\lambda}{\mathrm{d}\phi}+2\phi\lambda\bigg).
\end{equation}
In the following, we assume Eq. (\ref{eqlambda}), that is, $\lambda(\phi)=\lambda_2(\phi-\phi_0)^2$.

\subsection{Perturbation of the scalar field and the metric}
\label{pm}
We assume that the scalar field can be expanded in power series around its background value $\phi_0$ as \citet{Kopeikin2004} and define
\begin{equation}
	\label{scalarpert}
	\phi=\phi_0(1+\zeta),
\end{equation}
where $\zeta$ is dimensionless perturbation of the scalar field around $\phi_0$.

In particular, decomposition of the coupling function $\theta(\phi)$ can be written as
\begin{eqnarray}
	\label{scalardecomp}
	\theta(\phi) & = &\omega_0+\omega_1\zeta+\frac{1}{2}\omega_2\zeta^2+\dots,
\end{eqnarray}
where $\omega_0\equiv\theta(\phi_0)$ and $\omega_n\equiv(\mathrm{d}^n\theta/\mathrm{d}\zeta^n)_{\phi=\phi_0}$.

Following \citet{Chandra.1PN, Chandra.2PN}, we look for solutions of the field equations in the form of a Taylor expansion of the metric tensor and the scalar field with respect to the parameter $\varepsilon\equiv 1/c$ such that
\begin{eqnarray}
	\label{PPNmetric1}
	g_{00} & = & -1+\varepsilon^2\overset{(2)}{h}_{00}+\varepsilon^4\overset{(4)}{h}_{00}+\varepsilon^6\overset{(6)}{h}_{00}+\cdots,\\
	\label{PPNmetric2}
	g_{0i} & = & \varepsilon^3\overset{(3)}{h}_{0i}+\varepsilon^5\overset{(5)}{h}_{0i}+\cdots,\\
	\label{PPNmetric3}
	g_{ij} & = & \delta_{ij}+\varepsilon^2\overset{(2)}{h}_{ij}+\varepsilon^4\overset{(4)}{h}_{ij}+\cdots,\\
	\label{ezeta}\zeta & = & \varepsilon^2 \overset{(2)}{\zeta}+\varepsilon^4\overset{(4)}{\zeta}+\cdots,
\end{eqnarray}
and
\begin{eqnarray}
	\label{}
	T_{00} & = & \overset{(0)}{T_{00}}+\varepsilon^2\overset{(2)}{T_{00}}+\varepsilon^4\overset{(4)}{T_{00}}+\cdots,\\
	T_{0i} & = & \varepsilon\overset{(1)}{T_{0i}}+\varepsilon^3\overset{(3)}{T_{0i}}+\cdots,\\
	T_{ij} & = & \varepsilon^2\overset{(2)}{T_{ij}}+\varepsilon^4\overset{(4)}{T_{ij}}+\cdots.
\end{eqnarray}

Furthermore, we simplify the notations with the definitions:
\begin{equation}
	N\equiv\overset{(2)}{h}_{00}, \qquad L\equiv\overset{(4)}{h}_{00}\qquad L_i\equiv\overset{(3)}{h}_{0i} \qquad H_{ij} \equiv\overset{(2)}{h}_{ij}, \qquad H\equiv\overset{(2)}{h}_{kk},
\end{equation}
\begin{equation}
	\label{}
	Q\equiv\overset{(6)}{h_{00}},\qquad Q_i\equiv\overset{(5)}{h_{0i}}, \qquad Q_{ij}\equiv \overset{(4)}{h_{ij}}.
\end{equation}
According to Eqs. (\ref{scalarpert}), (\ref{scalardecomp}) and (\ref{ezeta}), we have
\begin{eqnarray}
	\label{theta(phi)}
	\theta(\phi) & = & \omega_0+\varepsilon^2\omega_1\overset{(2)}{\zeta}+\varepsilon^4\bigg(\frac{1}{2}\omega_2\overset{(2)}{\zeta}^2+\omega_1\overset{(4)}{\zeta}\bigg)+O(\varepsilon^6),\\
	\label{dthetadphi}
	\frac{\mathrm{d}\theta}{\mathrm{d}\phi} & = & \frac{1}{\phi_0}\bigg(\omega_1+\varepsilon^2\omega_2\overset{(2)}{\zeta}+O(\varepsilon^4)\bigg).
\end{eqnarray}
$O(\varepsilon^n)$ means of order $\varepsilon^n$. From here on, we omit the writing of $O(\varepsilon^n)$ where there is no ambiguity. We introduce three parameters $\gamma$, $\beta$ and $\iota$ as follows:
\begin{eqnarray}
	\label{gamma}
	\gamma & \equiv & \frac{\omega_0+1}{\omega_0+2},\\
	\label{beta}
	\beta & \equiv & 1+\frac{\omega_1}{(2\omega_0+3)(2\omega_0+4)^2},\\
	\label{iota}
	\iota & \equiv & \frac{1}{2}\frac{(\gamma-1)^4}{\gamma+1}\omega_2.
\end{eqnarray}
For scalar-tensor theory with $\lambda_2=0$, it turns out that $\gamma$ and $\beta$ become the values of corresponding standard PPN parameters $\gamma$ and $\beta$ in \citet{Will1972}.

Corresponding expansion of the energy-momentum tensor is 
\begin{eqnarray}
	\label{PPNTuv1}
	T_{00}&=&\rho^{\ast}+\varepsilon^2\rho^{\ast}\bigg[ \frac{1}{2}v^2+\Pi-N-\frac{1}{2}H\bigg],\nonumber\\
	&&\phantom{\rho}+\varepsilon^4\rho^{\ast}\bigg[\phantom{+}\frac{3}{8}v^4+\frac{1}{2}v^2\Pi-\frac{1}{4}v^2H+v^2\frac{p}{\rho^{\ast}}-\Pi N -\frac{1}{2}\Pi H-\frac{1}{2}N H\nonumber\\
	&&\phantom{\phantom{\rho}+\varepsilon^4\rho^{\ast}\bigg[}+\frac{1}{8}H^2+\frac{1}{4}H_{lk}H_{lk}+\frac{1}{2}H_{lk}v^lv^k -L-L_kv^k-\frac{1}{2}Q_{kk}\bigg],\\
	T_{0i}&=&-\varepsilon\rho^{\ast} v^i-\varepsilon^3\rho^{\ast}\bigg[v^i\bigg(\frac{1}{2}v^2+\Pi-\frac{1}{2}H+\frac{p}{\rho^{\ast}}\bigg)+L_i+H_{ik}v^k\bigg],\\
	\label{PPNTuv3}
	T_{ij}&=&\varepsilon^2(\rho^{\ast} v^i v^j +p\delta_{ij}).
\end{eqnarray}
Here we have used the invariant density $\rho^{\ast}\equiv\sqrt{-g}u^0\rho$ \citep{Fock} with its 2PN expression 
\begin{eqnarray}
	\label{}
	\rho^{\ast} & =  &\rho+\varepsilon^2\rho\bigg(\frac{1}{2}v^2+\frac{1}{2}H\bigg)\nonumber\\
	&&\phantom{\rho}+\varepsilon^4\rho\bigg[\phantom{+} \frac{3}{8}v^4+\frac{1}{2}v^2N+\frac{1}{4}v^2H+\frac{1}{2}H_{lk}v^lv^k\nonumber\\
	&&\phantom{\rho+\varepsilon^4\rho\bigg[}-\frac{1}{4}H_{lk}H_{lk}+\frac{1}{8}H^2+L_kv^k+\frac{1}{2}Q_{kk} \bigg].
\end{eqnarray}

\subsection{The gauge condition}

We use the gauge condition imposed on the component of the metric tensor proposed by \citet{Kopeikin2004} as follows:
\begin{equation}
	\label{KVgauge}
	\bigg(\frac{\phi}{\phi_0}\sqrt{-g}g^{\mu \nu}\bigg)_{,\nu}=0.
\end{equation}
Although it is called ``Nutku gauge'' by Kopeikin and Vlasov, the gauge condition is different from \citet{Nutku1969.1, Nutku1969.2}. To 2PN order, this gauge gives
\begin{eqnarray}
	\label{KNg1}
	&&\phantom{+} \varepsilon^2\bigg(\frac{1}{2}H_{,i}-\frac{1}{2}N_{,i}-H_{ik,k}+\overset{(2)}{\zeta}_{,i}\bigg)\nonumber\\
	&&+\varepsilon^4\bigg(-\frac{1}{2}NN_{,i}+\frac{1}{2}H_{ik}N_{,k}+H_{il}H_{lk,k}+H_{il,k}H_{Lk}-\frac{1}{2}H_{ik}H_{,k}\nonumber\\
	&&\phantom{+\varepsilon^4\bigg(}-\frac{1}{2}H_{lk}H_{lk,i}-\frac{1}{2}L_{,i}+L_{i,t}+\frac{1}{2}Q_{kk,i}-Q_{ik,k}-\frac{1}{2}\overset{(2)}{\zeta}N_{,i}\nonumber\\
	&&\phantom{+\varepsilon^4\bigg(}+\frac{1}{2}\overset{(2)}{\zeta}H_{,i}-H_{ik}\overset{(2)}{\zeta}_{,k}-\overset{(2)}{\zeta} H_{ik,k}+\overset{(4)}{\zeta}_{,i} \bigg)=0,
\end{eqnarray}
and
\begin{eqnarray}
	\label{KNg2}
	&&\phantom{+}\varepsilon^3\bigg(-\frac{1}{2}N_{,t}-\frac{1}{2}H_{,t}+L_{k,k}-\overset{(2)}{\zeta}_{,t}\bigg)\nonumber\\
	&&+\varepsilon^5\bigg(-\frac{1}{2}NH_{,t}-NN_{,t}+\frac{1}{2}H_{lk}H_{lk,t}-H_{lk}L_{l,k}+NL_{k,k}+\frac{1}{2}L_kN_{,k}\nonumber\\
	&&\phantom{+\varepsilon^5\bigg(}+\frac{1}{2}L_kH_{,k}-L_lH_{lk,k}-\frac{1}{2}L_{,t}+Q_{k,k}-\frac{1}{2}Q_{kk,t}\nonumber\\
	&&\phantom{+\varepsilon^5\bigg(}-\frac{1}{2}\overset{(2)}{\zeta}N_{,t}-N\overset{(2)}{\zeta}_{,t}-\frac{1}{2}\overset{(2)}{\zeta}H_{,t}+\overset{(2)}{\zeta}L_{k,k}+L_k\overset{(2)}{\zeta}_{,k}-\overset{(4)}{\zeta}_{,t}\bigg)=0.
\end{eqnarray}

\subsection{Metric coefficients}
With the metric, Eqs. (\ref{PPNmetric1})-(\ref{PPNmetric3}), the energy-momentum tensor, Eqs. (\ref{PPNTuv1})-(\ref{PPNTuv3}), the decomposition of scalar field, Eqs. (\ref{scalarpert}), (\ref{ezeta}), (\ref{theta(phi)}) and (\ref{dthetadphi}), the parameters, Eqs. (\ref{gamma})-(\ref{iota}) and the field equations (\ref{feg}) and (\ref{fep}), the metric coefficients can be solved. 

\subsubsection{Newtonian approximation}
The equation for $N$ is
\begin{equation}
	\label{}
	\nabla^2N=-8\pi G \rho^{\ast}-\xi_1 \overset{(2)}{\zeta},
\end{equation}
where $\overset{(2)}{\zeta}$ satisfies
\begin{equation}
	\label{}
	\nabla^2\overset{(2)}{\zeta}+\xi_1\overset{(2)}{\zeta}=-4(1-\gamma)\pi G \rho^{\ast},
\end{equation}
and
\begin{equation}
	\label{}
	\xi_1=4\frac{1-\gamma}{1+\gamma}\lambda_2\phi_0^2.
\end{equation}
The solution for $\overset{(2)}{\zeta}(\vec{r}\,)$ is
\begin{equation}
	\label{}
	\overset{(2)}{\zeta}(\vec{r}\,)=(1-\gamma)G\int_V \frac{\rho^{\ast}(\vec{r}\,')e^{-\xi_1 |\vec{r}-\vec{r}'|}}{|\vec{r}-\vec{r}\,'|}\mathrm{d}\vec{r}\,'.
\end{equation}
Hence, the solution for $N(\vec{r}\,)$ is
\begin{equation}
	\label{}
	N(\vec{r}\,)=2G\int_V \frac{\rho^{\ast}(\vec{r}\,')}{|\vec{r}-\vec{r}\,'|}\mathrm{d}\vec{r}\,'+\frac{\xi_1}{4\pi}\int_V \frac{\overset{(2)}{\zeta}(\vec{r}\,')}{|\vec{r}-\vec{r}\,'|}\mathrm{d}\vec{r}\,'.
\end{equation}
When $\lambda=0$, $\overset{(2)}{\zeta}$ and $N$ reduces to $(1-\gamma)U$ and $2U$, where
\begin{equation}
	\label{}
	U(\vec{r}\,)=\int_V  \frac{\rho^{\ast}(\vec{r}\,')}{|\vec{r}-\vec{r}\,'|}\mathrm{d}\vec{r}\,',
\end{equation}
is the Newtonian potential.

\subsubsection{1st post-Newtonian approximation}
Following the method developed by \citet{Chandra.1PN,Chandra.2PN} and \citet{Nutku1969.1}, we work out the 1st post-Newtonian approximation:
\begin{equation}
	\label{eqHij}
	\nabla^2 H_{ij} = \delta_{ij}(-8\gamma\pi G \rho^{\ast}+\xi_1\overset{(2)}{\zeta}),
\end{equation}
and
\begin{equation}
	\label{}
	\nabla^2 L_i = 8(\gamma+1) \pi G \rho^{\ast} v^i,
\end{equation}
where $G\equiv 2/[\phi_0(1+\gamma)]$. According to Eq. (\ref{eqHij}), we have $H_{ij}=V\delta_{ij}$.
For $L$, we have
\begin{eqnarray}
	\label{}
	\nabla^2 L & = & -8\pi G \rho^{\ast}\bigg[\frac{1}{2}(2\gamma+1)v^2-\frac{1}{2}V-N+\Pi+3\gamma\frac{p}{\rho^{\ast}}+\frac{4\beta-\gamma-3}{\gamma-1}\overset{(2)}{\zeta}\bigg]\nonumber\\
	&&-\frac{1}{2}N_{,k}N_{,k}-\frac{1}{2}N_{,k}V_{,k}+N_{,tt}-\overset{(2)}{\zeta}_{,k}N_{,k}-\frac{4(\beta-1)}{(\gamma-1)^2}\overset{(2)}{\zeta}_{,k}\overset{(2)}{\zeta}_{,k}\nonumber\\
	&&+\xi_1(N-V)\overset{(2)}{\zeta}-\xi_3\overset{(2)}{\zeta}^2-\xi_1\overset{(4)}{\zeta},
\end{eqnarray}
where
\begin{equation}
	\label{}
	\xi_3=4\frac{8\beta+\gamma-9}{\gamma^2-1}\lambda_2\phi_0^2.
\end{equation}
The Poisson equation for $\overset{(4)}{\zeta}$ is
\begin{eqnarray}
	\label{}
	\nabla^2 \overset{(4)}{\zeta} & = & -4\pi G \rho^{\ast} \bigg[\phantom{+}\frac{1}{2}(\gamma-1)v^2+\frac{1}{2}(\gamma-1)V -(\gamma-1)\Pi\nonumber\\
	&&\phantom{-4\pi G \rho^{\ast} \bigg[}+3(\gamma-1)\frac{p}{\rho^{\ast}}+\frac{8(\beta-1)}{\gamma-1}\overset{(2)}{\zeta}\bigg]\nonumber\\
	&&+\overset{(2)}{\zeta}_{,tt}+\frac{1}{2}\overset{(2)}{\zeta}_{,k}N_{,k}-\frac{1}{2}\overset{(2)}{\zeta}_{,k}V_{,k}+\frac{4(1-\beta)}{(\gamma-1)^2}\overset{(2)}{\zeta}_{,k}\overset{(2)}{\zeta}_{,k}\nonumber\\
	&&-\xi_1V\overset{(2)}{\zeta}+\xi_2\overset{(2)}{\zeta}^2,
\end{eqnarray}
with
\begin{equation}
	\label{}
	\xi_2=2\frac{3\gamma^2-6\gamma-16\beta+19}{\gamma^2-1}\lambda_2\phi_0^2.
\end{equation}
From these Poisson equations, the integrals can be readily written out.

\subsubsection{2nd post-Newtonian approximation}
As above, we solve for the 2nd post-Newtonian approximation:
\begin{eqnarray}
	\label{}
	\nabla^2 Q_{ij} & = & -8(1+\gamma) \pi G \rho^{\ast} v^i v^j\nonumber\\
	&&-\frac{1}{2}N_{,i}N_{,j}+VN_{,ij}-VV_{,ij}-\frac{1}{2}V_{,i}V_{,j}\nonumber\\
	&&-\overset{(2)}{\zeta}N_{,ij}-\frac{1}{2}\overset{(2)}{\zeta}_{,i}N_{,j}-\frac{1}{2}\overset{(2)}{\zeta_{,j}}N_{,i}+\frac{1}{2}\overset{(2)}{\zeta}_{,i}V_{,j}+\frac{1}{2}\overset{(2)}{\zeta}_{,j}V_{,i}\nonumber\\
	&&-2V\overset{(2)}{\zeta}_{,ij}+\overset{(2)}{\zeta}V_{,ij}+2\overset{(2)}{\zeta}\overset{(2)}{\zeta}_{,ij}+\frac{2(2\gamma-1)}{\gamma-1}\overset{(2)}{\zeta}_{,i}\overset{(2)}{\zeta}_{,j}\nonumber\\
	&&+\delta_{ij}\bigg\{ +8\pi G \rho^{\ast} \bigg[+\frac{1}{2}\gamma v^2+\frac{1}{2}\gamma V-\gamma\Pi+(2\gamma-1)\frac{p}{\rho^{\ast}} \nonumber\\
	&&\phantom{+\delta_{ij}\bigg\{ +8\pi G \rho^{\ast} \bigg[}+\frac{4\beta-4+\gamma^2-\gamma}{\gamma-1}\overset{(2)}{\zeta}\bigg]\nonumber\\
	&&\phantom{+\delta_{ij}\bigg\{}+\frac{1}{2}N_{,k}V_{,k}+\frac{1}{2}V_{,k}V_{,k}+V_{,tt}-\overset{(2)}{\zeta}_{,k}N_{,k}+\frac{4(\beta-1)}{(\gamma-1)^2}\overset{(2)}{\zeta}_{,k}\overset{(2)}{\zeta}_{,k}\nonumber\\
	&&\phantom{+\delta_{ij}\bigg\{}+VV_{,kk}+\xi_1V\overset{(2)}{\zeta}+\xi_3\overset{(2)}{\zeta}^2+\xi_1\overset{(4)}{\zeta}\bigg\},
\end{eqnarray}

\begin{eqnarray}
	\label{}
	\nabla^2 Q_i & = &\phantom{+} 8\pi G \rho^{\ast} \bigg[ (\gamma+1)v^i\bigg(\frac{1}{2}v^2+\Pi+\frac{1}{2}V+\frac{p}{\rho^{\ast}}-\overset{(2)}{\zeta}\bigg)+L_i \bigg]\nonumber\\
	&&+\frac{1}{2}N_{,i}N_{,t}+\frac{1}{2}NN_{,it}+N_{,i}V_{,t}+\frac{3}{2}NV_{,it}+\frac{1}{2}VN_{,it}-\frac{1}{2}V_{,i}V_{,t}\nonumber\\
	&&-\frac{1}{2}VV_{,it}-\frac{1}{2}N_{,k}L_{i,k}+\frac{1}{2}N_{,ik}L_k-N_{,k}L_{k,i}-NL_{k,ki}\nonumber\\
	&&-\frac{1}{2}V_{,ik}L_k+\frac{1}{2}V_{,k}L_{i,k}+V_{,i}L_{k,k}-V_{,k}L_{k,i}+L_{i,tt}\nonumber\\
	&&+\overset{(2)}{\zeta}_{,it}N-\frac{1}{2}\overset{(2)}{\zeta}_{,t}N_{,i}+\frac{1}{2}\overset{(2)}{\zeta}_{,i}N_{,t}-\overset{(2)}{\zeta}_{,it}V+2\overset{(2)}{\zeta}V_{,it}+\frac{3}{2}\overset{(2)}{\zeta}_{,i}V_{,t}\nonumber\\
	&&+\frac{1}{2}\overset{(2)}{\zeta}_{,t}V_{,i}-\overset{(2)}{\zeta}_{,i}L_{k,k}-\overset{(2)}{\zeta}L_{k,ki}-L_{i,k}\overset{(2)}{\zeta}_{,k}-\overset{(2)}{\zeta}_{,ik}L_k\nonumber\\
	&&+\frac{2(2\gamma-1)}{\gamma-1}\overset{(2)}{\zeta}_{,i}\overset{(2)}{\zeta}_{,t}+2\overset{(2)}{\zeta}\overset{(2)}{\zeta}_{,it}+\xi_1L_i\overset{(2)}{\zeta},
\end{eqnarray}
and
\begin{eqnarray}
	\label{}
	\nabla^2Q & = & -8\pi G \rho^{\ast} \bigg\{\phantom{+} \frac{1}{8}(4\gamma+3)v^4+\frac{1}{4}(2\gamma+1)v^2V+\frac{1}{2}(2\gamma+1)v^2\Pi\nonumber\\
	&&\phantom{-8\pi G \rho \bigg\{}+(\gamma+1)v^2\frac{p}{\rho^{\ast}}-\Pi N-\frac{1}{2}\Pi V-3\gamma N\frac{p}{\rho^{\ast}}+3\gamma V \frac{p}{\rho^{\ast}}\nonumber\\
	&&\phantom{-8\pi G \rho \bigg\{}+\frac{1}{2}NV+\frac{3}{8}V^2-L-L_kv^k-\frac{1}{2}Q_{kk}\nonumber\\
	&&\phantom{-8\pi G \rho \bigg\{}-\frac{2\gamma^2-\gamma+4\beta-5}{2(\gamma-1)}\overset{(2)}{\zeta}v^2+\frac{4\beta-\gamma-3}{\gamma-1}\overset{(2)}{\zeta}\bigg(\Pi-N-\frac{1}{2}V\bigg)\nonumber\\
	&&\phantom{-8\pi G \rho \bigg\{}+\bigg[1-\frac{\iota}{(\gamma-1)^2}-\frac{4(\beta-1)(\gamma^2+8\beta-2\gamma-7)}{(\gamma-1)^3}\bigg]\overset{(2)}{\zeta}^2\nonumber\\
	&&\phantom{-8\pi G \rho \bigg\{}-\frac{3(\gamma^2-\gamma+4\beta-4)}{\gamma-1}\overset{(2)}{\zeta}\frac{p}{\rho^{\ast}}+\frac{4\beta-\gamma-3}{\gamma-1}\overset{(4)}{\zeta}\bigg\}\nonumber\\
	&&+2NN_{,tt}-\frac{1}{2}NN_{,k}N_{,k}+2N_{,t}N_{,t}-\frac{1}{2}VN_{,k}V_{,k}+VN_{,tt}\nonumber\\
	&&+3NV_{,tt}+\frac{3}{2}N_{,t}V_{,t}+VN_{,k}N_{,k}-N_{,k}L_{,k}-L_{k,k}N_{,t}-2NL_{k,kt}\nonumber\\
	&&-N_{,k}L_{k,t}-L_kN_{,kt}+N_{,k}Q_{kl,l}+N_{,kl}Q_{kl}-\frac{1}{2}N_{,l}Q_{kk,l}-\frac{3}{2}V_{,t}V_{,t}\nonumber\\
	&&+L_{,tt}-\frac{1}{2}V_{,k}L_{,k}+2V_{,t}L_{k,k}-V_{,kt}L_k-L_{k,l}L_{l,k}+L_{l,k}L_{l,k}\nonumber\\
	&&+2\overset{(2)}{\zeta}_{,tt}N+\overset{(2)}{\zeta}N_{,tt}+2\overset{(2)}{\zeta}_{,t}N_{,t}+3\overset{(2)}{\zeta}_{,t}V_{,t}+3\overset{(2)}{\zeta}V_{,tt}-\overset{(2)}{\zeta}_{,k}L_{,k}\nonumber\\
	&&+2\overset{(2)}{\zeta}_{,k}L_{k,t}-2\overset{(2)}{\zeta}L_{k,kt}-2\overset{(2)}{\zeta}_{,t}L_{k,k}-2\overset{(2)}{\zeta}_{,kt}L_k-2\overset{(2)}{\zeta}_{,k}L_{,kt}\nonumber\\
	&&+\overset{(2)}{\zeta}\overset{(2)}{\zeta}_{,k}N_{,k}+2\overset{(2)}{\zeta}\overset{(2)}{\zeta}_{,tt}+\frac{2(2\gamma^2-3\gamma+2\beta-1)}{(\gamma-1)^2}\overset{(2)}{\zeta}_{,t}\overset{(2)}{\zeta}_{,t}\nonumber\\
	&&+\frac{4(\beta-1)}{(\gamma-1)^2}N\overset{(2)}{\zeta}_{,k}\overset{(2)}{\zeta}_{,k}+\frac{2\iota}{(\gamma-1)^3}\overset{(2)}{\zeta}\overset{(2)}{\zeta}_{,k}\overset{(2)}{\zeta}_{,k}-\overset{(4)}{\zeta}_{,k}N_{,k}\nonumber\\
	&&+\frac{4(\beta-1)(\gamma^2+8\beta-2\gamma-7)}{(\gamma-1)^4}\overset{(2)}{\zeta}\overset{(2)}{\zeta}_{,k}\overset{(2)}{\zeta}_{,k}-\frac{8(\beta-1)}{(\gamma-1)^2}\overset{(2)}{\zeta}_{,k}\overset{(4)}{\zeta}_{,k}\nonumber\\
	&&+\xi_1(VN+L)\overset{(2)}{\zeta}-2\xi_3\overset{(2)}{\zeta}\overset{(4)}{\zeta}+\xi_3(N-V)\overset{(2)}{\zeta}^2\nonumber\\
	&&-\xi_6\overset{(2)}{\zeta}^3+\xi_1(N-V)\overset{(4)}{\zeta}-\xi_1\overset{(6)}{\zeta},
\end{eqnarray}
where $\overset{(6)}{\zeta}$ satisfies
\begin{eqnarray}
	\label{}
	\nabla^2\overset{(6)}{\zeta}+\xi_1\overset{(6)}{\zeta} & = & -4\pi G \rho^{\ast} \bigg\{+\frac{\gamma-1}{8}v^4+\frac{\gamma-1}{2}v^2\Pi+\frac{\gamma-1}{4}v^2(2N+V)\nonumber\\
	&&\phantom{-4\pi G \rho^{\ast} \bigg\{}-\frac{4(\beta-1)}{\gamma-1}v^2\overset{(2)}{\zeta}+\frac{\gamma-1}{2}\Pi V+\frac{8(\beta-1)}{\gamma-1}\Pi\overset{(2)}{\zeta}\nonumber\\
	&&\phantom{-4\pi G \rho^{\ast} \bigg\{}+3(\gamma-1)V\frac{p}{\rho^{\ast}}-\frac{24(\beta-1)}{\gamma-1}\frac{p}{\rho^{\ast}}\overset{(2)}{\zeta}-\frac{3}{8}(\gamma-1)V^2\nonumber\\
	&&\phantom{-4\pi G \rho^{\ast} \bigg\{}+(\gamma-1)\bigg(L_kv^k+\frac{1}{2}Q_{kk}\bigg)-\frac{4(\beta-1)}{\gamma-1}V\overset{(2)}{\zeta}\nonumber\\
	&&\phantom{-4\pi G \rho^{\ast} \bigg\{}-\frac{2(32\beta^2+\iota\gamma-64\gamma-\iota+32)}{(\gamma-1)^3}\overset{(2)}{\zeta}^2+\frac{8(\beta-1)}{\gamma-1}\overset{(4)}{\zeta}\bigg\}\nonumber\\
	&&+N\overset{(2)}{\zeta}_{,tt}+\frac{1}{2}\overset{(2)}{\zeta}_{,t}N_{,t}+\frac{3}{2}\overset{(2)}{\zeta}V_{,t}+\frac{1}{2}N\overset{(2)}{\zeta}_{,k}N_{,k}+\frac{1}{2}V\overset{(2)}{\zeta}_{,k}V_{,k}\nonumber\\
	&&-2\overset{(2)}{\zeta}_{,kt}L_k+\frac{1}{2}\overset{(2)}{\zeta}_{,k}L_{,k}-\overset{(2)}{\zeta}_{,t}L_{k,k}-\overset{(2)}{\zeta}_{,k}L_{k,t}+\overset{(2)}{\zeta}_{,k}Q_{kl,l}\nonumber\\
	&&+\overset{(2)}{\zeta}_{,lk}Q_{lk}-\frac{1}{2}\overset{(2)}{\zeta}_{,l}Q_{kk,l}+V\overset{(2)}{\zeta}_{,tt}+\frac{4(\beta-1)}{(\gamma-1)^2}\overset{(2)}{\zeta}_{,t}\overset{(2)}{\zeta}_{,t}\nonumber\\
	&&+\frac{32(\beta-1)^2+2\iota(\gamma-1)}{(\gamma-1)^4}\overset{(2)}{\zeta}\overset{(2)}{\zeta}_{,k}\overset{(2)}{\zeta}_{,k}-\frac{8(\beta-1)}{(\gamma-1)^2}\overset{(2)}{\zeta}_{,k}\overset{(4)}{\zeta}_{,k}\nonumber\\
	&&+\frac{1}{2}\overset{(4)}{\zeta}_{,k}N_{,k}-\frac{1}{2}\overset{(4)}{\zeta}_{,k}V_{,k}+\overset{(4)}{\zeta}_{,tt}+\xi_2V\overset{(2)}{\zeta}^2+\xi_4\overset{(2)}{\zeta}^3\nonumber\\
	&&+\xi_5\overset{(2)}{\zeta}\overset{(4)}{\zeta}-\xi_1V\overset{(4)}{\zeta},
\end{eqnarray}
with
\begin{eqnarray}
	\label{}
	\xi_4 & = & \frac{2\lambda_2\phi_0^2}{(\gamma-1)^3(\gamma+1)}\bigg(\gamma^4-4\gamma^3-24\beta\gamma^2+30\gamma^2+128\beta^2+48\beta\gamma\nonumber\\
	&&\phantom{\frac{2\lambda_2\phi_0^2}{(\gamma-1)^3(\gamma+1)}\bigg(}+4\iota\gamma-52\gamma-4\iota-280\beta+153\bigg),
\end{eqnarray}
\begin{equation}
	\label{}
	\xi_5=\frac{4(3\gamma^3-6\gamma-16\beta+19)}{\gamma^2-1}\lambda_2\phi_0^2,
\end{equation}
and
\begin{equation}
	\label{}
	\xi_6=\frac{8(2\beta\gamma^2-2\gamma^2-32\beta^2-4\beta\gamma-\iota\gamma+\iota+4\gamma+66\beta-34)}{(\gamma-1)^3(\gamma+1)}.
\end{equation}

\subsubsection{Summary of the parameters}
The physical meaning of $\lambda_2$ is that it gives an inverse range $\xi_1$ $(=4(1-\gamma)\lambda_2\phi_0^2/(1+\gamma))$ of the intermediate gravity. With $\lambda_2\ne0$ the theory violates the inverse square law for gravitation \citep{Fischbach,Li}. When $\lambda_2=0$, $\gamma$ and $\beta$ reduce to the standard PPN parameters in 1st post-Newtonian approximation. It is clear that, besides them, only one parameter $\iota$ emerges in the 2nd post-Newtonian approximation, which represents the 3rd order nonlinearity in $g_{00}$. Table \ref{tab1} gives a summary of the parameters involved and their values in general relativity. 
\begin{table}
	\centering
	\caption{\label{tab1}Summary of the parameters.}
	\begin{tabular}{c|l|c|l}
		\hline
		Parameter & What it measures, relative to GR & Value in GR & Value in STT\\
		\hline
		$\gamma$ & How much space curvature $(g_{ij})$ & 1 & $\frac{\omega_0+1}{\omega_0+2}$\\
		&  is produced by unit rest mass?& & \\
		&  \citep{MTW} & & \\
		$\beta$ & How much 2nd order nonlinearity is there  & 1 & $1+\frac{\omega_1}{(2\omega_0+3)(2\omega_0+4)^2}$\\
		& in the superposition law for gravity $(g_{00})$? & & \\
		&  \citep{MTW} & & \\
		\hline
		$\iota$ & How much 3rd order nonlinearity is there  & 0 & $\frac{\omega_2}{2(3+2\omega_0)(\omega_0+2)^3}$\\
		& in the superposition law for gravity $(g_{00})$?&  &\\
		\hline
	 \end{tabular}
\end{table}

In 1992, Damour and Esposito-Far\`{e}se proposed a multiscalar-tensor theory \citep{Damour1992}. When only one scalar field involved, the action reads,
\begin{equation}
	\label{del}
	S_{\ast} = \frac{c^3}{16\pi G_{\ast}}\int \mathrm{d}^4x\sqrt{g_{\ast}}[R_{\ast}-2g_{\ast}^{\mu \nu}\partial_{\mu}\varphi\partial_{\nu}\varphi]+S_m[\psi_m,\tilde{g}_{\mu\nu}]
\end{equation}
where $\tilde{g}_{\mu \nu}$ and $g^{\ast}_{\mu \nu}$ are the physical metric and Einstein-frame metric respectively, and $\tilde{g}_{\mu \nu}=A^2(\varphi)g^{\ast}_{\mu\nu}$. Hereafter, tilde will be dropped for clarity. After setting
\begin{equation}
	\label{}
	\phi\equiv\frac{1}{A^2G_{\ast}},
\end{equation}
and
\begin{equation}
	\label{}
	\theta(\phi)\equiv-\frac{3}{2}+\frac{1}{2}\bigg(\frac{\mathrm{d}\ln A}{\mathrm{d}\phi}\bigg)^{-2},
\end{equation}
the action (\ref{del}) could reduce to ours (\ref{lgst2}) in the case that $\lambda_2\equiv0$. In a following paper \citep{Damour1996}, Damour and Esposito-Far\`{e}se derived 2PN approximation of their theory, and showed that it would introduce only two new 2PN parameters $\varepsilon_{DE}$ and $\zeta_{DE}$. (Here we use the subscript or superscript ``DE'' to denote Damour and Esposito-Far\`{e}se's results.) When there is only one scalar-field involved, $\zeta_{DE}$, depending only on $\bar{\gamma}\equiv\gamma-1$ and $\bar{\beta}\equiv\beta-1$, is \emph{not} a new parameter in 2PN; $\varepsilon_{DE}$ is the only new independent parameter in 2PN related to our parameter $\iota$ by $\varepsilon_{DE}=-\iota-12\bar{\beta}^2/\bar{\gamma}$ (see Table \ref{tab2}).
\begin{table}
	\centering
	\caption{\label{tab2}A parameters' comparison between Damour and Esposito-Far\`{e}se's (DE's) and ours when only one scalar field involved, where $\bar{\gamma}\equiv\gamma-1$, $\bar{\beta}\equiv\beta-1$, $\alpha_0={\partial \ln A(\varphi_0)}/{\partial \varphi_0}$, $\beta_0={\partial \alpha(\varphi_0)}/{\partial \varphi_0}$, $\beta_0'={\partial \beta(\varphi_0)}/{\partial \varphi_0}$ (see \citet{Damour1996} for details).}
	\begin{tabular}{c|c|l|l}
	\hline
	Order (PN) & Paramters & DE's & Ours\\
	\hline
	1 & $\bar{\gamma}$ & $-\frac{2\alpha_0^2}{1+\alpha_0^2}$ & $-\frac{1}{\omega_0+2}$\\
	1 & $\bar{\beta}$  & $\frac{1}{2}\frac{\beta_0\alpha_0^2}{(1+\alpha_0^2)^2}$ & $\frac{\omega_1}{4(2\omega_0+3)(\omega_0+2)^2}$\\
	2 & $\varepsilon_{DE}$ & $\frac{\beta_0'\alpha_0^3}{(1+\alpha_0^2)^3}$ & $-\iota-\frac{12\bar{\beta}^2}{\bar{\gamma}}$\\
	2 & $\zeta_{DE}$ & $\frac{\beta_0^2\alpha_0^2}{(1+\alpha_0^2)^3}$ & $-\frac{8\bar{\beta}^2}{\bar{\gamma}}$\\
	\hline
	\end{tabular}
\end{table}
Damour and Esposito-Far\`{e}se derived the 2PN deviation from GR of the metric component $g_{00}$, but not other metric components. Their 2PN deviation from GR $\delta g^{DE}_{00}$ is given by
\begin{equation}
	\label{dDE}
	\nabla^2\delta g^{DE}_{00}= \frac{\varepsilon_{DE}}{3c^6}\nabla^2U^3-\frac{\varepsilon_{DE}}{c^6}4\pi G \sigma U^2+O\bigg(\frac{\bar{\gamma}}{c^6},\frac{\bar{\beta}}{c^6}\bigg)+O\bigg(\frac{1}{c^8}\bigg).
\end{equation}
Our corresponding equation is
\begin{equation}
	\label{dst}
	\nabla^2\delta g_{00}= -\frac{\iota}{c^6}U\nabla^2U^2+O\bigg(\frac{\bar{\gamma}}{c^6},\frac{\bar{\beta}}{c^6}\bigg)+O\bigg(\frac{1}{c^8}\bigg).
\end{equation}
With the help of
\begin{equation}
	\label{}
	\nabla^2U=-4\pi G \sigma,
\end{equation}
and
\begin{equation}
	\label{}
	\frac{1}{3}\nabla^2U^3=U\nabla^2U^2-U^2\nabla^2U,
\end{equation}
Eqs. (\ref{dDE}) and (\ref{dst}) can be transformed into each other and agree. This is a consistency check for own calculation.

\section{Equations of Motion}
\label{sec2}
The equations of motion for perfect fluid are derived in 2nd post-Newtonian approximation.

$T^{i\nu}_{;\nu}=0$ yields the momentum equation,
\begin{eqnarray}
	\label{}
	&&\bigg(\frac{\mathrm{d}}{\mathrm{d}t}+\nabla\cdot\vec{v}\bigg)\bigg\{\phantom{+}\rho^{\ast} v^i+\varepsilon^2\rho^{\ast} v^i \bigg[\frac{1}{2}v^2+\Pi+N-\frac{3}{2}V+\frac{p}{\rho^{\ast}}\bigg]\nonumber\\
	&&\phantom{\bigg(\frac{\mathrm{d}}{\mathrm{d}t}+\nabla\cdot\vec{v}\bigg)\bigg\{}+\varepsilon^4\rho^{\ast} v^i\bigg[\phantom{+}\frac{3}{8}v^4+\frac{1}{2}v^2\Pi+v^2\frac{p}{\rho^{\ast}}+v^2\bigg(N-\frac{1}{4}V\bigg)\nonumber\\
	&&\phantom{\bigg(\frac{\mathrm{d}}{\mathrm{d}t}+\nabla\cdot\vec{v}\bigg)\bigg\{+\varepsilon^4\rho^{\ast} v^i\bigg[} +\Pi \bigg(N-\frac{3}{2}V\bigg)+\frac{p}{\rho^{\ast}}N+\frac{15}{8}V^2\nonumber\\
	&&\phantom{\bigg(\frac{\mathrm{d}}{\mathrm{d}t}+\nabla\cdot\vec{v}\bigg)\bigg\{+\varepsilon^4\rho^{\ast} v^i\bigg[} -\frac{3}{2}NV+N^2+L+L_kv^k-\frac{1}{2}Q_{kk}\bigg]\bigg\}\nonumber\\
	&&+p_{,i}-\frac{1}{2}\rho N_{,i}\nonumber\\
	&&+\varepsilon^2\rho^{\ast}\bigg\{\phantom{+}\frac{1}{2}v^i\bigg(5V_{,t}-N_{,t}+5V_{,k}v^k-N_{,k}v^k\bigg)\nonumber\\
	&&\phantom{+\varepsilon^2\rho^{\ast}\bigg[} -N_{,i}\bigg(\frac{1}{4}v^2+\frac{1}{2}\Pi+\frac{1}{2}\frac{p}{\rho^{\ast}}+\frac{1}{2}N-\frac{5}{4}V\bigg)-\frac{1}{2}V_{,i}v^2\nonumber\\
	&&\phantom{+\varepsilon^2\rho^{\ast}\bigg[}-\frac{V}{\rho^{\ast}}p_{,i}+L_{i,t}-\frac{1}{2}L_{,i}+L_{i,k}v^k-L_{k,i}v^k\bigg\}\nonumber\\
	&&+\varepsilon^4\bigg(L_{i}p_{,t}+V^2p_{,i}-p_{,k}Q_{ik}\bigg)\nonumber\\
	&&+\varepsilon^4p\bigg\{\phantom{+}\frac{5}{2}v^i\frac{\mathrm{d}}{\mathrm{d}t}V-\frac{1}{2}v^i\frac{\mathrm{d}}{\mathrm{d}t}N+\frac{\mathrm{d}}{\mathrm{d}t}L_i-\frac{1}{2}L_{,i}-L_{k,i}v^k\nonumber\\
	&&\phantom{+\varepsilon^4p\bigg[}-\frac{1}{2}N_{,i}(N+v^2)+\frac{1}{2}V_{,i}(N-v^2)\bigg\}\nonumber\\
	&&+\varepsilon^4\rho^{\ast}\bigg\{\phantom{+}v^i\bigg[\phantom{+}\frac{1}{2}\frac{\mathrm{d}}{\mathrm{d}t}(Q_{kk}-L)-\frac{1}{2}\bigg(v^2+\Pi+2N-\frac{3}{2}V\bigg)\frac{\mathrm{d}}{\mathrm{d}t}N\nonumber\\
	&&\phantom{+\varepsilon^4\rho^{\ast}\bigg\{\phantom{+}v^i\bigg[}+\frac{5}{2}\bigg(v^2+\Pi+N-\frac{5}{2}V\bigg)\frac{\mathrm{d}}{\mathrm{d}t}V+\frac{1}{4}v^2\bigg(\frac{\mathrm{d}}{\mathrm{d}t}N-5\frac{\mathrm{d}}{\mathrm{d}t}V\bigg)\bigg]\nonumber\\
	&&\phantom{+\varepsilon^4\rho^{\ast}\bigg[}-N_{,i}\bigg(\phantom{+}\frac{3}{16}v^4+\frac{1}{4}v^2\Pi+\frac{1}{2}v^2N-\frac{3}{8}v^2V+\frac{1}{2}\Pi N\nonumber\\
	&&\phantom{+\varepsilon^4\rho^{\ast}\bigg[-U_{,i}\bigg(}-\frac{5}{4}\Pi V-\frac{1}{2}\frac{p}{\rho^{\ast}}V+\frac{1}{2}N^2-\frac{5}{4}NV+\frac{35}{16}V^2\nonumber\\
	&&\phantom{+\varepsilon^4\rho^{\ast}\bigg[-U_{,i}\bigg(}+\frac{1}{2}L+\frac{1}{2}L_kv^k-\frac{1}{4}Q_{kk}\big)]\nonumber\\
	&&\phantom{+\varepsilon^4\rho^{\ast}\bigg[}-V_{,i}\bigg(\frac{1}{4}v^4+\frac{1}{2}v^2N-\frac{5}{4}v^2V+\frac{1}{2}v^2\Pi+\frac{1}{2}\frac{p}{\rho^{\ast}}N\bigg)\nonumber\\
	&&\phantom{+\varepsilon^4\rho^{\ast}\bigg\{}+L_i\bigg(\frac{1}{2}\frac{\mathrm{d}}{\mathrm{d}t}N+\frac{1}{2}N_{,k}v^k\bigg)+\bigg(v^2+\Pi+N-V\bigg)\frac{\mathrm{d}}{\mathrm{d}t}L_i\nonumber\\
	&&\phantom{+\varepsilon^4\rho^{\ast}\bigg\{}-\bigg(L_{k,i}v^k+\frac{1}{2}L_{,i}\bigg)\bigg(\frac{1}{2}v^2+\Pi+N-\frac{5}{2}V\bigg)\nonumber\\
	&&\phantom{+\varepsilon^4\rho^{\ast}\bigg\{}-(L_{i,k}v^k+L_{i,t})\bigg(\frac{1}{2}v^2+\frac{3}{2}N\bigg)+\frac{\mathrm{d}}{\mathrm{d}t}Q_i-\frac{1}{2}Q_{,i}\nonumber\\
	&&\phantom{+\varepsilon^4\rho^{\ast}\bigg[}+v^k\frac{\mathrm{d}}{\mathrm{d}t}Q_{ik}-Q_{k,i}v^k+\frac{1}{2}Q_{ik}N_{,k}-\frac{1}{2}Q_{kl,i}v^kv^l\bigg\}=0.
\end{eqnarray}

$T^{0\nu}_{;\nu}=0$ gives the continuity equation,
\begin{eqnarray}
	\label{}
	&& \bigg(\frac{\mathrm{d}}{\mathrm{d}t}+\nabla\cdot\vec{v}\bigg)\bigg\{\phantom{+}\rho^{\ast}+\varepsilon^2\rho^{\ast}\bigg[\frac{1}{2}v^2+\Pi+N-\frac{3}{2}V\bigg]\nonumber\\
	&&\phantom{\bigg(\frac{\mathrm{d}}{\mathrm{d}t}+\nabla\cdot\vec{v}\bigg)\bigg\{}+\varepsilon^4\rho^{\ast}\bigg[\phantom{+}\frac{3}{8}v^4+\frac{1}{2}v^2\Pi+v^2\bigg(N-\frac{1}{4}V\bigg)+v^2\frac{p}{\rho^{\ast}}\nonumber\\
	&&\phantom{\bigg(\frac{\mathrm{d}}{\mathrm{d}t}+\nabla\cdot\vec{v}\bigg)\bigg\{+\varepsilon^4\rho\bigg(}+\Pi\bigg(N-\frac{3}{2}V\bigg)+\frac{15}{8}V^2-\frac{3}{2}NV+N^2\nonumber\\
	&&\phantom{\bigg(\frac{\mathrm{d}}{\mathrm{d}t}+\nabla\cdot\vec{v}\bigg)\bigg\{+\varepsilon^4\rho\bigg(}  +L+2L_kv^k-\frac{1}{2}Q_{kk}\bigg]\bigg\}\nonumber\\
	&&+\varepsilon^2\bigg\{\nabla\cdot(p\vec{v}) + \rho^{\ast}\bigg(\frac{3}{2}\frac{\mathrm{d}}{\mathrm{d}t}V-\frac{\mathrm{d}}{\mathrm{d}t}N-\frac{1}{2}N_{,k}v^k\bigg)\bigg\}\nonumber\\
	&&+\varepsilon^4\bigg\{\phantom{+}\nabla\cdot(pN\vec{v})+L_kp_{,k}\nonumber\\
	&&\phantom{+\varepsilon^4\bigg\{}+\rho^{\ast}\bigg[-\bigg(+\frac{1}{2}v^2+\Pi-\frac{3}{2}V+2N\bigg)\frac{\mathrm{d}}{\mathrm{d}t}N\nonumber\\
	&&\phantom{+\varepsilon^4\bigg\{+\rho^{\ast}\bigg[}+\bigg(+\frac{3}{4}v^2+\frac{3}{2}\Pi+\frac{3}{2}N-\frac{15}{4}V\bigg)\frac{\mathrm{d}}{\mathrm{d}t}V\nonumber\\
	&&\phantom{+\varepsilon^4\bigg\{+\rho^{\ast}\bigg[}+\frac{1}{2}\frac{\mathrm{d}}{\mathrm{d}t}(Q_{kk}-L)+\frac{1}{2}V_{,t}\bigg(v^2+3\frac{p}{\rho^{\ast}}\bigg)\nonumber\\
	&&\phantom{+\varepsilon^4\bigg\{+\rho^{\ast}\bigg[} -N_{,k}v^k\bigg(\frac{1}{4}v^2+\frac{1}{2}\Pi+\frac{3}{2}\frac{p}{\rho^{\ast}}+N-\frac{3}{4}V\bigg)\nonumber\\
	&&\phantom{+\varepsilon^4\bigg\{+\rho^{\ast}\bigg[}+\frac{3}{2}V_{,k}v^k\frac{p}{\rho^{\ast}}-\frac{1}{2}L_{,k}v^k-L_{k}U_{,k}-L_{k,l}v^kv^l\bigg]\bigg\}=0.
\end{eqnarray}

\section{Conservation Laws}
\label{sec3}
We adopt the energy-momentum complex to obtain the conservation laws in a simple way and we set $\lambda_2=0$ for simplicity. The energy-momentum complex $\Theta^{\mu \nu}$ is defined as in \cite{Nutku1969.2}
\begin{equation}
	\label{}
	\Theta^{\mu \nu}=\phi_0^{-1}\phi(-g)(T^{\mu \mu}+t^{\mu \nu}),
\end{equation}
where
\begin{equation}
	\label{}
	t^{\mu \nu}=\frac{\phi}{\varepsilon^28\pi}\bigg[\frac{1}{2(-g)\phi^2}U^{\mu \alpha \nu \beta}_{,\alpha \beta}-X^{\mu \nu}\bigg],
\end{equation}
in which
\begin{equation}
	\label{}
	X^{\mu \nu}=R^{\mu \nu}-\frac{1}{2}g^{\mu \nu}R-\frac{\theta(\phi)}{\phi^2}\bigg(\phi^{,\mu}\phi^{,\nu}-\frac{1}{2}g^{\mu \nu}\phi_{,\sigma}\phi^{,\sigma}\bigg)-\frac{1}{\phi}\bigg(\phi^{;\mu \nu}-g^{\mu \nu}\Box_g\phi\bigg),
\end{equation}
and
\begin{equation}
	\label{}
	U^{\mu \alpha \nu \beta}=\phi^2(-g)(g^{\mu\nu}g^{\alpha\beta}-g^{\mu \beta}g^{\nu \alpha}).
\end{equation}
Kopeikin-Nutku gauge Eq. (\ref{KVgauge}) gives
\begin{equation}
	\label{}
	U^{\mu \alpha \nu \beta}_{,\alpha \beta}=\phi\sqrt{-g}g^{\alpha \beta}(\phi\sqrt{-g}g^{\mu \nu})_{,\alpha \beta}-(\phi\sqrt{-g}g^{\mu \beta})_{,\alpha}(\phi\sqrt{-g}g^{\nu \alpha})_{,\beta}.
\end{equation}
To obtain the order we need, we extend the metric expansion of $g_{\mu \nu}$ as
\begin{eqnarray}
	\label{}
	g_{00} & = & -1+2\varepsilon^2U+\varepsilon^4L+\varepsilon^6Q,\\
	g_{0i} & = & \varepsilon^3L_i+\varepsilon^5Q_i,\\
	\label{sij}
	g_{ij} & = & \delta_{ij}+2\varepsilon^2\gamma\delta_{ij}U+\varepsilon^4Q_{ij}+\varepsilon^6S_{ij},
\end{eqnarray}
with $S_{ij}$ defined by Eq. (\ref{sij}).

Thus, we can obtain conserved quantities: 
\begin{enumerate}
\item energy:
\begin{equation}
	\label{}
	E=c^2\int (\Theta^{00}-\rho u^0 \sqrt{-g}) \mathrm{d}^3x;
\end{equation}

\item linear momentum:
\begin{equation}
	\label{}
	P^i=\int \Theta^{0i}\mathrm{d}^3x;
\end{equation}

\item angular momentum:
\begin{equation}
	\label{}
	L_{i}=\varepsilon_{ikl}\int \Theta^{0k}x_l \mathrm{d}^3x.
\end{equation}
\end{enumerate}
Modulo divergence \citep{Chandra.complex}, $\Theta^{00}$ and $\Theta^{0i}$ can be expressed as
\begin{eqnarray}
	\label{}
	\Theta^{00} &=&\phantom{+}\rho^{\ast}+\varepsilon^2\rho^{\ast}\bigg(\frac{1}{2}v^2+\Pi-\frac{1}{2}U\bigg)\nonumber\\
	&&+\varepsilon^4\rho^{\ast}\bigg[\phantom{+}\frac{3}{8}v^4+\bigg(3\gamma+\frac{9}{4}\bigg)v^2U+\frac{1}{2}v^2\Pi+(4\gamma-1)\Pi U \nonumber\\
	&&\phantom{+\varepsilon^4\rho^{\ast}\bigg(}+\frac{1}{2}\bigg(26\gamma^2+11\gamma+32\beta-37+\frac{32}{\gamma+1}\bigg)U^2\nonumber\\
	&&\phantom{+\varepsilon^4\rho^{\ast}\bigg(} +\frac{1}{2}L_kv^k+\frac{p}{\rho}v^2-4(2\gamma-1)\frac{p}{\rho^{\ast}}U\bigg]\nonumber\\
	&&+\varepsilon^4 \frac{1}{4\pi G}\bigg[\phantom{+}2\bigg(3\gamma^2-10\gamma-3\beta+7-\frac{16}{\gamma+1} \bigg)UU_{,k}U_{,k}\nonumber\\
	&&\phantom{+\varepsilon^4 \frac{1}{4\pi G}\bigg(}+\frac{1}{2}U_{,t}U_{,t}-4\gamma UU_{,tt}\bigg].
\end{eqnarray}
and
\begin{equation}
	\label{}
	\frac{1}{\varepsilon}\Theta^{0i}=\pi_i+\theta_{ik,k},
\end{equation}
where
\begin{eqnarray}
	\label{}
	\pi_i&=&\phantom{+}\rho^{\ast} v^i+\varepsilon^2\rho^{\ast} v^i \bigg(\frac{1}{2}v^2+\Pi-U+\frac{p}{\rho^{\ast}}\bigg)-\varepsilon^2\frac{1}{4\pi G}U_{,i}U_{,t}\nonumber\\
	&&+\varepsilon^4\rho^{\ast}\bigg\{\phantom{+}v^i\bigg[\phantom{+}\frac{3}{8}v^4+\bigg(\gamma+\frac{1}{2}\bigg)v^2U+\frac{1}{2}v^2\Pi-\frac{1}{2}(2\gamma+1)^2U^2\nonumber\\
	&&\phantom{+\varepsilon^4\rho\bigg\{\phantom{+}v^i\bigg[}-U\Pi+(3\gamma-1)\frac{p}{\rho^{\ast}}U+\frac{p}{\rho^{\ast}}v^2\nonumber\\
	&&\phantom{+\varepsilon^4\rho\bigg\{\phantom{+}v^i\bigg[}-\frac{1}{2}L+L_kv^k+\frac{1}{2}Q_{kk}+\overset{(4)}{\zeta}\bigg]\nonumber\\
	&&\phantom{+\varepsilon^4\rho\bigg\{}+L_i\bigg[-\frac{1}{2}v^2+\frac{1-\gamma}{2(1+\gamma)}\Pi+\frac{2\gamma-1}{\gamma+1}\frac{p}{\rho^{\ast}}\nonumber\\
	&&\phantom{+\varepsilon^4\rho\bigg\{+L_i\bigg[} +\frac{1}{2}\bigg(8-9\gamma+\frac{4(1-\beta)}{1-\gamma}-\frac{8\beta}{\gamma+1}\bigg)U\bigg]\nonumber\\
	&&\phantom{+\varepsilon^4\rho\bigg\{}+Q_{ik}v^k\bigg\}\nonumber\\
	&&+\varepsilon^4\frac{1}{8\pi G}\bigg[-2\bigg( 11\gamma^2-6\gamma+4\beta+11-\frac{16}{\gamma+1}\bigg)UU_{,i}U_{,t}+\frac{2}{\gamma+1}U_{,t}L_{,i}\nonumber\\
	&&\phantom{+\varepsilon^4\frac{1}{8\pi G}\bigg[}-\frac{2}{\gamma+1}U_{,i}L_{,t}+\frac{4(\gamma-1)}{\gamma+1}U_{,t}L_{i,t}+\frac{\gamma-3}{\gamma+1}L_iU_{,tt}\nonumber\\
	&&\phantom{+\varepsilon^4\frac{1}{8\pi G}\bigg[}-\frac{\gamma+5}{\gamma+1}UL_{i,tt}+(6\gamma-3)L_iU_{,k}U_{,k}-\frac{4}{\gamma+1}UL_kU_{,ki}\nonumber\\
	&&\phantom{+\varepsilon^4\frac{1}{8\pi G}\bigg[}-\frac{4(2\gamma^2+\gamma-2)}{\gamma+1}UU_{,k}L_{k,i}-\frac{2(\beta-1)(\gamma-3)}{1-\gamma^2}L_iU_{,k}U_{,k}\nonumber\\
	&&\phantom{+\varepsilon^4\frac{1}{8\pi G}\bigg[}-\frac{2}{\gamma+1}L_{i,k}L_{k,t}+\frac{1}{\gamma+1}L_kL_{,ki}-\frac{1}{\gamma+1}Q_{i,tt}\nonumber\\
	&&\phantom{+\varepsilon^4\frac{1}{8\pi G}\bigg[}-\frac{\gamma-1}{\gamma+1}U_{,k}Q_{k,i}-\frac{2(\gamma-1)}{\gamma+1}U_{,i}Q_{k,k}-\frac{3\gamma-1}{\gamma+1}UQ_{ik,kt}\nonumber\\
	&&\phantom{+\varepsilon^4\frac{1}{8\pi G}\bigg[}-\frac{2\gamma}{\gamma+1}Q_{ik,t}U_{,k}+\frac{2}{\gamma+1}Q_{ik}U_{,kt}-\frac{2(2\gamma-1)}{\gamma+1}U_{,t}Q_{ik,k}\nonumber\\
	&&\phantom{+\varepsilon^4\frac{1}{8\pi G}\bigg[}-\frac{\gamma-1}{\gamma+1}Q_{kk}U_{,it}+\frac{1}{\gamma+1}L_{i,l}Q_{lk,k}-\frac{1}{\gamma+1}L_lQ_{lk,ki}\nonumber\\
	&&\phantom{+\varepsilon^4\frac{1}{8\pi G}\bigg[}-\frac{1}{\gamma+1}S_{kk,it}+\frac{1}{\gamma+1}S_{ik,kt}-\frac{2(\gamma-1)}{\gamma+1}\overset{(4)}{\zeta}U_{,it}\bigg],
\end{eqnarray}
and
\begin{eqnarray}
	\label{}
	\theta_{ik}&=&\phantom{+}\varepsilon^2\frac{1}{4\pi G}\bigg[2(\gamma+1)\delta_{ik}UU_{,t}+UL_{i,k}-UL_{k,i}\bigg]\nonumber\\
	&&+\varepsilon^4\frac{1}{8\pi G}\bigg[-\frac{4\gamma(3\gamma^2-4\gamma+1)}{\gamma+1}\delta_{ik}U^2U_{,t}-\frac{15\gamma^2-14\gamma+3}{2(\gamma+1)}U^2L_{i,k}\nonumber\\
	&&\phantom{+\varepsilon^4\frac{1}{8\pi G}\bigg[}-\frac{(3\gamma-1)^2}{(\gamma+1)}UL_iU_{,k}-\frac{2(3\gamma-5)}{\gamma+1}\delta_{ik}UL_lU_{,l}-\frac{1}{2(\gamma+1)}L_iL_{,k}\nonumber\\
	&&\phantom{+\varepsilon^4\frac{1}{8\pi G}\bigg[}+\frac{1}{\gamma+1}(L_lQ_{ik,l}-Q_{il}L_{k,l}+L_iQ_{kl,l}-L_{l,i}Q_{kl}+L_lQ_{li,k})\nonumber\\
	&&\phantom{+\varepsilon^4\frac{1}{8\pi G}\bigg[}-\frac{3\gamma-1}{\gamma+1}UQ_{k,i}+\frac{2}{\gamma+1}U_{,i}Q_k+\frac{3\gamma-1}{\gamma+1}\delta_{ik}(U_{,t}Q_{ll}-UQ_{ll,t})\nonumber\\
	&&\phantom{+\varepsilon^4\frac{1}{8\pi G}\bigg[}-\frac{1}{2(\gamma+1)}LL_{i,k}+\frac{2(\gamma-1)}{\gamma+1}Q_iU_{,k}+\frac{4\gamma}{\gamma+1}UQ_{i,k}\bigg].
\end{eqnarray}

\section{Solar System Dynamics and Test of Scalar-Tensor Theory}
\label{sec4}
To test the scalar-tensor theory including intermediate-range gravitational force in the solar system, we need equations of motion for solar-system dynamics.

Since intermediate-range gravitational force has not been discovered in the solar system, it must be small. The solution of $\overset{(2)}{\zeta}$ has a $(1-\gamma)$ factor. The intermediate range part of $N$ has a $(1-\gamma)^2$ factor. From the constraint of empirical test \citep{Bertotti2003}, $(1-\gamma)$ should be less than $10^{-4}$. Hence the deviation from Newtonian gravity should be less than $10^{-8}$ and $\lambda_2$ is not much constrained. Further experiments/observations with better precision will be able to measure $\lambda_2$ better. From the metric, we can obtain the geodesic equation of motion:
\begin{equation}
	\label{}
	\frac{\mathrm{d}^2\vec{r}}{\mathrm{d}t^2}=\vec{a}_N+\vec{a}_{1PN}+\vec{a}_{2PN},
\end{equation}
where $a_N$ is the Newtonian acceleration and it has the form
\begin{equation}
	\label{}
	\vec{a}_N=\frac{1}{2}\nabla N =-G\int_V \frac{\rho^{\ast}(\vec{r}\,')(\vec{r}-\vec{r}\,')}{|\vec{r}-\vec{r}\,'|^3}\mathrm{d}\vec{r}\,'-\frac{\xi_1}{8\pi}\int_V \frac{\overset{(2)}{\zeta}(\vec{r}\,')(\vec{r}-\vec{r}\,')}{|\vec{r}-\vec{r}\,'|^3}\mathrm{d}\vec{r}\,'.
\end{equation}
The full equation of motion will be given and discussed in \cite{Dong}.

\section{Outlook}
\label{sec5}
Full 2PN approximation of general scalar-tensor theory of gravity has been obtained in a single frame. Multiple-frame studies like those have been worked out for general relativity \citep{DSX1,DSX2,DSX3}, for PPN formalism with two parameters \citep{Klioner2000}, and for scalar-tensor theory \citep{Kopeikin2004} in the 1PN approximation would be the next step to investigate. 2PN approximation of vector-tensor theory of gravity \citep{Jacobson2001,Luo} and other theories need to be worked out to see a more general structure of 2PN approximation. With this done, it would be easier to formulate a useful parameterized 2PN formalism for testing 2nd order relativistic gravity.

This research is supported by the National Natural Science Foundation of China under Grants No. 10475114 (W.-T. Ni) and No. 10563001 (Y. Xie and T.-Y. Huang) and the Foundation of Minor Planets of Purple Mountain Observatory. We thank Prof. R. Caldwell for pointing out a typo.



\end{document}